\documentclass[11pt]{article}

\usepackage[margin=1in]{geometry}
\usepackage{amsmath,amssymb,amsthm}
\usepackage{hyperref}
\usepackage{graphicx}
\usepackage{enumitem}

\usepackage[T1]{fontenc}
\usepackage[utf8]{inputenc}
\usepackage{lmodern}

\title{What is the AGI in Offensive Security?}
\author{
  Youngwoong Cho \\
  \texttt{ycho14@sheffield.ac.uk}
}
\date{}

\begin{document}

\maketitle

\begin{abstract}
What is the AGI in Offensive Security? One can break it down into two questions : (1) any offensive security tasks could be reduced into symbolic language manipulation (language representation + reasoning), (2) powerful language model (LLM) are enough to "deal with" any symbolic language manipulation. This paper can formally model a target system as a state machine and a hacker as an interactive symbolic agent. And it shows that every interaction in an offensive engagement can be encoded as a finite string. This paper provides definitions, short lemmas, and open discussion. 
\end{abstract}

\section{Introduction}
Hacking is often seen as a dark art combining deep level analysing skill and creative thinking. However, at its core, any offensive security task on a computer system is fundamentally a computational process. This paper proposes "any offensive security (hacking) tasks can be reduced to symbolic language manipulation," given the assumption that target systems are digital, algorithmically computable, and their inputs or outputs can be distinct symbols. 

This paper does not assume the reader already agrees that hacking is "just code (if so, the problem was solved already)" - instead, this paper will demonstrate it by constructing a model and deriving the claim. This paper begins by modeling a computer system as a state machine, which is a standard mathematical abstraction. And this paper formalizes the hacker as an interactive policy or agent that supplies inputs to the target system and observes outputs. With these models which captures the essence of the process, this paper show how any interaction can be encoded as string (a sequence of symbols) and how any strategy the hacker employs can be seen as a function producing next according to given ones. Building on these insights, this paper presents the possibility of this suggestion "if these (models and assumptions) truly captures the true and essential process of it, LLM (or more realistically MLLM) could be AGI of Hacking."

Even though the paper explicitly mentioned "LLM" multiple times, the main message of the paper is not that "LLM could be AGI of Hacking." What the paper really is trying to answer is "What is hacking, structurally ?"

\section{Modeling Target Systems as State Machines}
\label{sec:state-machines}

A state machine is a mathematical framework for modelling a system by specifying a collection of states and the transitions that move the system from one state to another. Each state reflects a specific condition of the system, and transitions represent how the system changes in response to inputs or events. State machines can be finite (having a limited number of states) or even infinite. Real computer systems often have an astronomically large (practically infinite) number of possible states due to variables, memory, etc. Formally, this paper can use the notion of a transition system to allow potentially infinite states while still treating the system in state-transition terms.

Definition 1 (Target System as State Machine). A digital target system is modelled as a state machine (or transition system) M=(S, $\Sigma$, O, $g$) where:

- S is the set of states (potentially infinite, but countable, representing all possible configurations of the system).
    
- $\Sigma$ is the input alphabet, a finite set of symbols from which input strings are built (e.g. the set of all bytes 0–255, or ASCII characters).
    
- O is the output alphabet, a finite set of symbols for outputs (possibly the same as $\Sigma$, e.g. bytes).
    
- $g$:S×$\Sigma$*→S×O* is the (possibly partial) transition function or I/O function, which takes the current state and an input string and produces a new state and an output string (the system’s response). This paper allow $g$ to operate on input sequences so that multi-step interactions can be composed; alternatively, one could define a stepwise transition $g'$:S×$\Sigma$→S×O and extend it to sequences.

This definition is essentially an abstract description of a computing system. It may help to relate it to a real example: imagine S includes all contents of memory and disk, the values of registers, etc., for a given program. The input alphabet $\Sigma$ could be bytes sent over a network connection to this program, and the output alphabet O could be bytes the program sends back. The transition function $g$ describes how the program processes inputs and changes its internal state and outputs accordingly. Because this paper assumes the system is computable (it follows definite algorithmic rules), such a $g$ exists in principle and can be thought of as implemented by the program’s code. In fact, by the Church-Turing Thesis, any discrete, algorithmic process can be simulated by a Turing machine – which itself is a kind of state machine. Thus, any digital computer system can be represented in this state machine framework. This modelling is standard in theoretical computer science and formal methods: for example, model-checking tools convert software into a state-transition model and then search for certain state paths.

It is important to note that in this model, the system’s inputs and outputs are discrete symbols. This is justified by the nature of digital computers: IO = sequences of 0s and 1s. All information – be it text, binary executables, network packets, etc. – is encoded in bits, and bits are discrete symbols (binary digits). Even high-level constructs like user commands or API calls ultimately reduce to machine-level bit patterns. Thus, assuming inputs/outputs are discreditable is not a stretch; it is an empirical reality of digital systems. One can treat any input as a string over some finite alphabet (often binary {0,1}, or for human readability, perhaps ASCII characters). Likewise, any output the system produces is a string over some finite alphabet.

 
Example (State Machine View of a Login System): Consider a simple server that has a login prompt. It might be modeled with states \[
S = \{\text{WAIT\_FOR\_USERNAME},\ \text{WAIT\_FOR\_PASSWORD},\ \text{LOGGED\_IN},\ \text{LOGGED\_OUT}, \ldots\}
\] The input alphabet $\Sigma$ could be the set of all possible keyboard characters a user can type. The output alphabet O could be the set of characters it can display. The transition function $g$ will move the system from WAITFORUSERNAME to WAITFORPASSWORD when a username string ending with newline is input, and so on. If a correct password is input, the state transitions to LOGGEDIN and outputs a welcome message; a wrong password might transition to WAITFORPASSWORD again (with a warning output). And it showed LOGGEDOUT after the user logout his or her auth. This is a deterministic example; more complex systems might have nondeterministic behaviour, but that can be handled in the model by having $g$ yield sets of possible next states instead of a single state.


\section{Modelling Hackers as Interactive Symbolic Agents}
\label{sec:hackers-agents}

Having modelled the target, this paper turn to the hacker or attacker. In offensive security scenarios, an attacker is an entity that interacts with the target system by providing inputs (payloads, commands, packets, etc.) and observing outputs (responses, side-channel signals, etc.), all with the aim of reaching some goal state (typically one that is advantageous to the attacker and not intended by the system’s designers).

This paper model the hacker abstractly as an interactive agent – essentially, a program or policy that can adapt its inputs based on what it has seen from the system so far. This interactive aspect is crucial. Unlike a one-shot input-output scenario, hacking often involves multiple steps: probe the system, see the reaction, then choose the next action, and so forth. Traditional Turing machines consume all input at once and then produce output, but here this paper need a model of ongoing interaction. In theoretical computer science, this is sometimes called an interactive Turing machine. As van Leeuwen and Wiedermann describe, an interactive Turing machine is an extension of the classical model that “allows new, unforeseen inputs to appear as the computation proceeds, instead of all inputs being fixed before the computation starts”. This matches the real-world hacking: the attacker can supply new inputs in reaction to the system’s outputs at each step.

 

Definition 2 (Hacker as Policy/Agent). A hacker agent (or attacker policy) is an algorithmic procedure that, given the history of the interaction with the system so far, produces the next input to send. Formally, this paper can represent the attacker as a (partial) function or strategy $\pi$:$(O^*)^*$→$\Sigma$*. The domain $(O^*)^*$ represents all possible finite sequences of output strings (essentially, the history of outputs from the system across multiple steps), and the domain $\Sigma$* is the set of possible input strings the attacker can send next. If the attacker also has internal memory or state, that can be modelled as part of the input to $\pi$ (by including the attacker’s state in the history, or by allowing $\pi$ to be stateful). For simplicity, one can imagine $\pi$ works like this: it reads the transcript of the interaction so far (everything the system has output up to now, perhaps including a record of what inputs were sent), and then it computes a new input string to send in. This $\pi$ is effectively the “mind” of the hacker. The key is: $\pi$ is symbolic, meaning it operates by reading and writing symbols (just like any computer program). The hacker’s decision-making is just an information-processing procedure.

 

This definition frames the hacker as an interactive function that keeps taking the system’s outputs and producing new inputs. Note that $\pi$ could be randomized or adaptive; those nuances can be added by letting $\pi$ make random choices or by expanding the notion of state to include probabilistic behaviour. But even a randomized strategy can be modelled as a distribution over deterministic functions, so conceptually it’s still within the realm of computation.

 

Example (Attacker Policy): In the login system example, an attacker’s strategy $\pi$ might be: always input “admin” as the username, then try a list of common passwords one by one. This strategy can be encoded as a simple algorithm: it ignores the system’s output until maybe it sees either “Welcome” (success) or “Locked out” (failure). If it sees a password prompt (some output indicating it’s waiting for a password), it sends the next password from its list. This $\pi$ can be implemented as a short program or even described in pseudocode. More sophisticated $\pi$ could adapt based on subtle outputs (e.g., timing differences or error messages) – those too are just functions of the observed output history.

 
At this point, this paper suggests two interacting machines: the target system M and the attacker policy $\pi$. When this paper composes them (connecting the output of one to the input of the other and vice versa), they form a combined dynamical system. One can imagine a loop: the attacker $\pi$ feeds an input to M, M processes it and transitions state, producing some output, the attacker $\pi$ reads that output (updating its own state or history) and produces another input, and so on. This dance continues until either the attacker stops or the system terminates or resets. Importantly, all information exchanged in this dance is symbolic (any IO - strings going back and forth)

\section{Encoding Interactions and Strategies as Strings}
\label{sec:encoding-strings}

Having established that both the target system and the attacker operate on symbolic inputs/outputs, this paper now argues that any interaction history can itself be represented as a string. This is a small and important milestone: if one can encode an entire hacking scenario (or transcript) as a single string, then reasoning about hacks becomes reasoning about strings.

 
Lemma 1.1 (Interaction Transcript Encoding). Any finite interaction (sequence of inputs and outputs exchanged) between the attacker and the system can be represented as finite sequence of symbols.

More precisely, 
$$T = (i_1, o_1, i_2, o_2, \dots, i_n, o_n), \quad \text{where } i_k \in \Sigma^*; o_k \in O^*$$
or for history h 
$$h = (i_1, o_1, i_2, o_2, \dots, i_n, o_n)$$
Then there exists a finite alphabet $\Gamma$ and an injective function

$$\mathrm{enc} : (\Sigma^* \times O^*)^* \to \Gamma^*$$

such that $\mathrm{enc}(T)$ uniquely represents $T$ as a set of symbols.

Proof Sketch. 
Because $\Sigma$ and $O$ are finite alphabets, $\Sigma^*$ and $O^*$ are countable sets of finite strings. (For finite interactions), we can encode pairs and sequences of finite strings into a set of symbols(string) by using standard delimiter-based or self-delimiting encodings (This is the easiest way, but just for proving that it is possible). 

Define: 

- $\mathrm{enc}_\Sigma(i)$: copy $i$ symbol-by-symbol into the $\Sigma$-copy inside $\Gamma$
    
- $\mathrm{enc}_O(o)$: copy $o$ symbol-by-symbol into the $O$-copy inside $\Gamma$
and then set any delimiter symbols such as (\#, \$)

Then, the finite sequence of interaction can be represented as
$$\mathrm{enc}(T) = \# \mathrm{enc}_\Sigma(i_1) \$ \mathrm{enc}_O(o_1) \# \mathrm{enc}_\Sigma(i_2) \$ \mathrm{enc}_O(o_2) \dots \# \mathrm{enc}_\Sigma(i_n) \$ \mathrm{enc}_O(o_n)$$

So any finite interactive exchange can be expressed into a set of symbols. 

Lemma 1.2 (Countable Symbols Sets admit tokenization for Sequence Models). 
Let $X$ be any countable set of abstract symbols. Then there exists an injective encoding 
$$\phi : X \to V^*$$
into finite strings over some finite vocabulary $V$. Consequently, any finite sequence over $X$ can be represented as a finite token sequence over $V$, making it a valid object for sequence-model processing. 

Proof Sketch. 
Since $X$ is countable, so $f: X \to \mathbb{N}$. And every natural number has a finite representation over a finite alphabet (even binary digits are possible, because every natural number has a finite binary representation). More generally, there exists a finite vocabulary $V$ and a computable, injective encoding $\phi : X \to V^*$

For example, one may take $V = \{0, 1, \mathtt{sep}\}$. Define $\phi(x)$ as the binary representation of $f(x)$ followed by a separator token $\mathtt{sep}$. This produces an injective mapping that assigns each abstract symbol to a finite token string. 

Extend $\phi$ to finite sequences by concatenation:

$$\phi(x_1, x_2, \dots, x_m) = \phi(x_1)\phi(x_2)\cdots\phi(x_m) \in V^*.$$

Thus, any finite sequence over $X$ can be serialized into a token sequence over finite vocabulary $V$. Consequently, at the level of input representation, any finite sequence over $X$ can be made compatible with token-based sequence models (e.g. LLM architectures). 

Remark. This lemma mainly states that symbol domains can be represented as token sequences.

Lemma 1.1 says one can flatten the time-sequence of hacking into a document-like form (a set of symbols). Imagine reading a log of the attack T. This is important because (according to any digital computer modelling) the true essence of hacking are inside the realm of strings.

 
Next, this paper considers the strategy $\pi$ itself. The attacker’s strategy or policy is a function (or an algorithm, which can be represented as code). This paper claim that the strategy, being a computable procedure, can also be encoded as a string and more importantly dealt with as a string essentially. In fact, it’s a well-known fact from computer science that any algorithm can be encoded in a finite description (think of a program’s source code or binary). For example, one can encode an entire Turing machine’s description as a finite string. In practice, an exploit script or program is literally a file of bits, i.e. a string. But even if the attacker is a human “running” a mental algorithm, this paper can in theory write down a description of what steps the human is following.

Lemma 2.1 (Strategy Encoding).  An attacker’s policy $\pi$, understood as an effective (computable) procedure that maps finite interaction histories (composed of input and output symbol sequences) to subsequent input symbols, admits a finite description as a string over some fixed alphabet $\Upsilon$ (for example, ASCII characters or binary digits). Equivalently, if $\pi$ is computable, then there exists a finite string $P \in \Upsilon^*$ that encodes the full logic of $\pi$ in some chosen representation scheme (such as a program, circuit description, or formal specification).

Proof Sketch. A computable policy is an algorithm, and every algorithm has a finite description. 
By definition, $\pi$ decides its next output by a finite, effectively executable procedure (e.g. algorithm, decision tree, or script).
Classical computability theory tells that any algorithm can be represented as a finite sequence of symbols e.g. a Turing machine description, more realistically and simply a program in a chosen language. 
So $\pi$ has a canonical, finite symbolic representation. By selecting a fixed alphabet $\Upsilon$, we can encode that representation as a single finite string ($P \in \Upsilon^*$). (On these settings, we don't need to care about the specific representation) Therefore, the entire logic of $\pi$ fits inside a finite symbolic artifact as claimed.

Lemma 2.2 (Strategy Learnability via LLM). Let $\pi$ be a computable attacker policy as defined above. 
Because $\pi$ induces a computable mapping from symbolic histories to symbolic next-moves, there exists a conditional sequence model-specifically, a large language model with suitable parameters ($\theta$)-whose output distribution approximates the action selection behaviour of $\pi$.    

More formally, for history $h$, an LLM approximate:
$$\Pr_\theta(x \mid h) \approx \delta_{\pi(h)}(x),$$
where $\delta_{\pi(h)}(x)$ denotes a point distribution centred at $\pi(h)$.

Proof Sketch. $\pi$ defines a next-token function over symbol sequences and LLMs can approximate conditional distributions over symbol sequences; the structures align. 
From prior lemmas, histories $(O^*)^*$ and outputs $\Sigma^*$ (from the attacker's perspective this is the output) lie in a countable symbolic domain that can be tokenized and processed by an LLM. 
 $\pi$ induces a computable conditional mapping. For each history $h$, the policy produces a next input $\pi(h)$. 
 This induces the deterministic conditional distribution over strings:
 $$\Pr(x \mid h) = \begin{cases} 1 &\text{if } x = \pi(h),\\[4pt] 0 &\text{otherwise}. \end{cases}$$

Any computable symbolic distribution can be approximated by large enough sequence model. 

LLM has the capacity to approximate distributions over token sequences (for countable vocabularies). 

Therefore, there exist parameters $\theta$ such that: 
$$\Pr_\theta(x \mid h) \to \delta_{\pi(h)}(x)$$
(in the limit of sufficient data, capacity, and training - realistically)
So the behavioural rule of $\pi$ is within the representational capacity of a language model. 
Because $\pi$ operates entirely over sequences of symbols, an LLM can in principle be trained to approximate its decision function. 

Remark (Program-Level Learnability). Because $\pi$ admits a finite symbolic encoding ($P \in \Upsilon^*$) (Lemma 2.1), an LLM may also learn to generate such an encoding (at a higher layer). 
This corresponds not to behavioural imitation of $\pi$'s action-decision, but to program synthesis, i.e., producing a symbolic description whose execution implements $\pi$. 
This capability is distinct from the approximation claim of Lemma 2.2, but compatible with the broader observation that symbolic policies lie within the expressive regime of LLMs.

Taken together, Lemma 1.1, Lemma 1.2, Lemma 2.1, and Lemma 2.2 yield two structural conclusions.
First, any offensive security task can be represented entirely in terms of symbolic language objects (system descriptions, inputs, outputs, and transcripts over suitable alphabets), and such symbolic objects can be tokenized into finite sequences over a fixed vocabulary, making them representationally compatible with LLM-style sequence models (Lemma 1.2). 
Second, a successful attack consists in following a computable policy mapping interaction histories to strategically chosen next inputs; Lemma 2.1 guarantees such a policy admits a finite symbolic encoding, and Lemma 2.2 guarantees that its induced behavioural mapping can, in principle, be approximated by a large language-model architecture.

Formally, consider a hacking scenario with three symbolic components:

(1) a description of the target system (source code, protocol specification, or a state-machine model);

(2) a description of the attacker’s policy, encoded as a finite procedure; and

(3) an execution trace recording their interaction.

All such components are finite strings over finite alphabets.

Let $\Sigma$ denote the input alphabet accepted by the target system, and let $L_{\text{exploit}} \subseteq \Sigma^{*}$ be the set of input sequences driving the system from its initial state into a designated goal state (what attacker wants).

The attacker’s objective is to produce some word $w \in L_{\text{exploit}}$.

Under the symbolic-encoding lemmas, both $\Sigma^{*}$ and $L_{\text{exploit}}$ are language-theoretic objects, and the attacker’s policy is a computable function from histories to inputs.

Under the symbol-processing and strategy-learnability lemmas, such a policy is, in principle, amenable to approximation by a large language model.

In this sense, hacking (or at least the true essence of it) is actually language representation + language-guided reasoning, and large language models are naturally suited to operate them. 

\section{Hacking Strategies as Symbolic String-to-String Procedures}
\label{sec:string-procedures}

This section brings together the earlier lemmas to state the central claim: offensive security activity can be viewed entirely as symbolic processing, and strategies of that kind fall, in principle, within the modelling scope of large language model architectures.

Informal Theorem(Hacking as Symbolic Sequential Computation).  
Under the assumptions that digital systems are discrete, computable, and ultimately operate over finite sequences of symbols, any offensive security task can be modeled as a reasoning problem of manipulating or crafting strings over finite alphabets. The attacker’s strategy can be characterized as a computable procedure it read the symbolic record of what has happened so far and decides which input string to send next. The task of finding a successful attack reduces to finding one or more input strings that, when issued according to such a procedure, drive the system into a designated target state. In light of Lemma 1.1 and Lemma 1.2, interaction histories and system behaviours admit finite symbolic encodings, and in light of Lemma 2.1 and Lemma 2.2, a computable attacker policy admits both a finite symbolic description and a behavioural mapping that is, at least in principle, approximable by a conditional sequence model. 

The earlier lemmas establish each ingredient of this statement. Lemma 1.1 says that everything exchanged during an attack-the entire transcript of prompts, commands, responses, error messages, and so on-can be represented as a sequence of symbols drawn from some alphabet. Lemma 1.2 asserts that any countable set of symbols admits a tokenization into finite sequences over a fixed vocabulary, so interaction artifacts can be represented in the input space of LLM-style architectures.
On the adversarial side, Lemma 2.1 observes that an effective attacker operates according to a policy that is itself computable; that policy can be encoded as a finite description string over some alphabet (for example, an abstract program in source code or byte form). Finally, Lemma 2.2 observes that because the attacker's policy defines a function from symbolic histories to symbolic inputs, and both the domain and range of this mapping are sequences of symbols, the resulting behaviour can be cast as a conditional distribution over next inputs given past transcripts, of exactly the LLM aims to approximate.
Since the attacker's policy admits a symbolic encoding, an LLM may not only approximate the behavioural mapping but may also, at a higher representational layer, learn to generate such an encoding. This corresponds to program synthesis rather than behavioural imitation. 

Seen from this perspective, THE CORE PART OF HACKING (IO between the target system and a hacker) is nothing more than a symbolic procedure that consumes strings and emits strings. Initially, the attacker possesses some prior knowledge about the target system e.g. IP addresses, service identifiers, documentation, public source code, results from earlier reconnaissance. All of that prior knowledge can be serialized into one or more strings. The attacker then applies a procedure—perhaps implemented in a human mind, a script, a toolchain, or some combination—to compute one or more candidate exploit inputs, again representable as strings. 
Interactive attacks simply iterate this process: the procedure reads the latest system response, updates its internal state, and emits the next input. However, since Lemma 1.1 guarantees that the entire interaction history can be represented as a finite string, and Lemma 2.1 ensures that the procedure is computable, the whole process can be treated as a single effective transformation from an initial symbolic description into a final attack transcript. 
In this view, the attacker is an agent performing symbolic transformations on symbolic inputs with reasoning power. 

An important consequence of this view is that no step in the offensive workflow escapes the realm of computation or language. The system being attacked is itself a program executing instructions over discrete memory and registers. The protocol that governs communication is defined in terms of finite bit patterns and message formats. The vulnerability conditions—buffer overflow, logic flaws, authentication bypasses, etc—can be formalized as predicates over program states or execution traces, all of which admit encodings as strings. 
The attacker’s reasoning process, to the extent that it yields an effective method for choosing inputs, can be represented as an algorithm and thus as a finite description string, as stated in Lemma 2.1. 
Thus, solving a hacking problem is structurally similar to solving a formal puzzle: one searches a symbolic space for a string satisfying certain constraints, where both the constraints and the candidate solutions are represented as symbolic entities.

Automated exploit generation provides a concrete illustration of this symbolic perspective. In such systems, exploit discovery is explicitly formulated as a search problem over symbolic representations of program behaviour. The target program is converted into a set of logical constraints that characterize when an exploited state occurs, for example when control flow is redirected to an attacker-controlled program counter. In these systems, program inputs are viewed as strings, and any parts that are not yet known are simply marked as symbolic variables. The solver looks for input strings that meet every condition needed to reach the specified goal state. Throughout this process, the constraints, symbolic variables, intermediate representations, and final exploit inputs are all encoded as strings in a formal language. The entire pipeline-from program parsing and formula construction through intermediate representations, constraint solving, and synthesizing exploits-functions by systematically transforming strings into new strings under specified rules. This is exactly the kind of symbolic string manipulation captured by the informal theorem and supporting lemmas.

Introducing large language models into this setting does not change the underlying structure. Logs, system messages, vulnerability notes, tool outputs, and exploit attempts are all just strings that can be tokenized within a language model’s vocabulary, as mentioned by the symbol-processing lemma. The attacker's policy corresponds to a computable function that maps past interaction histories to subsequent inputs, thereby inducing a conditional distribution over strings. That mapping can be expressed as a conditional probability distribution over strings, of the kind that a language-model-style architecture is designed to approximate, as captured by the strategy learnability lemma. By training the model on a sufficiently large corpus of interaction traces, it learns the pattern that maps each history to the next likely input. Thus, offensive security tasks become instances of symbolic sequence-modelling : given a symbolic interaction history, generate the next sequence of symbols that moves the overall state closer to an adversarial goal.

\section{Conclusion}
\label{sec:conclusion}

Under the strong but explicit assumptions of a digital, discrete, and computable environment, offensive security tasks—hacking in everyday terms—admit a reduction to symbolic language manipulation. The formal model presented in this perspective treats target systems as state machines and attackers as interactive symbolic policies, with their interplay captured entirely by finite strings. System behaviour, interaction transcripts, and attacker strategies all admit encodings into symbolic sequences. The goal of compromising a system then becomes the task of identifying one or more input strings that, when interpreted by the system’s transition rules, lead to an unintended or security-violating state. Within this framework, offensive techniques manifest as computable operations on symbolic representations, reinforcing the view that hacking is language-reducible and resides within the same conceptual space as formal languages, logic, and computability theory.

This viewpoint offers a unifying interpretation of offensive activity: the attacker is effectively programming the target with inputs, constructing an unintended program over the system’s input language. Any successful subversion corresponds to one more program—one more string—that happens to drive execution into a harmful configuration. Recognizing this brings powerful theoretical tools into scope, including ideas from computability, complexity, and formal language theory, for reasoning about what kinds of attacks are expressible, detectable, or preventable. At the same time, the vastness of the space of possible strings and the undecidability of many semantic properties point out why security is intrinsically difficult: the security problem is computation in an open, adversarial regime.

Seeing offensive security in this way also suggests a research direction for the role of large language models. When interactions, system responses, and attacker strategies are represented as strings, and attacker policies operate as effective procedures on those strings, language-model-style architectures become natural candidates for approximating such policies. This does not imply that current models are sufficient for fully autonomous offensive capabilities, nor that they can perfectly capture all relevant reasoning. Instead, it indicates that offensive tasks sit inside the broader class of symbolic sequence problems that these architectures are designed to handle in principle. Formal modelling of systems and interactions—through attack graphs, state exploration, or symbolic encodings of protocols—can therefore be integrated with learned sequence models within a rigorous, scientific security methodology. 

Taken together, these observations support a conceptual shift: hacking is structured symbolic manipulation. Every system accepts inputs drawn from a particular "language," and vulnerabilities correspond to dangerous regions within those languages. Strengthening security then becomes a matter of understanding and constraining the languages systems recognize, identifying harmful strings, and ensuring they cannot reach sensitive states or put it in the proper contexts. And this is the natural area of LLM. They can operate directly on the same symbolic structures-protocols, transcripts, programs, and inputs-and possess the capacity to navigate, generate, and reason over these languages with a flexibility previously unique to human experts. It suggests the potential that this general-purpose engines actually can be used for both attack and defence. 

\section{Open Discussion}
\label{sec:open-discussion}

Scope of Assumptions: This paper assumed all relevant systems are digital and computable. Consequently, the framework does not attempt to model human cognition, analog physical processes, or the full complexity of hardware and world-interacting systems. This does not mean that the core idea is irrelevant to those domains; rather, it reflects the fact that their fundamental signals e.g. physical world-are not natively language-friendly, even if we can describe them using our own language. On handling those tasks, an LLM is not, by itself, a complete solution. Still, for offensive security on those systems (essentially symbolic interface), the main idea may be applied. 

MLLM (more realistical perspective).
Multimodal large language models expand the width of the interaction history that an attacker can interact with. The history interaction or O may include screenshots, memory dumps, GUI states, or more sensory formats. At least, in reality, this significantly increases the usability and applicability of the model. 
Despite this extension, it is important to note that the core of hacking remains symbolic manipulation. Whether reasoning over text logs, binary blobs, network packets, or visual displays, the attacker's computable strategy $\pi$ acts on tokenized representations and emits new token sequences as candidate inputs. 

The Future.
Looking forward, the evolution of multimodal LLMs hints at a future where these models operate as a kind of Artificial General Intelligence for security work. As these models gain the ability to integrate text, code, binary artifacts, visual interfaces, and execution traces into one symbolic space, they gain the flexibility needed to understand and (possibly) reason about every aspect of an attack or defence workflow. Rather than acting as narrow assistants, they can become broad problem-solvers that can explore exploit patterns, synthesize policies, generate algorithms, and identify structural weaknesses across diverse modalities. In this sense, continued advances in multimodal LLMs suggest the new coming present human analysts stop doing the symbolic reasoning alone, need to collaborate with intelligent models.

\end{document}